\documentclass[aps,prb,reprint,showpacs,superscriptaddress,groupedaddress]{revtex4-1}
\usepackage{graphicx}
\usepackage{amsmath}
\usepackage{amssymb}
\usepackage{dcolumn}
\usepackage{dsfont}
\usepackage{latexsym}
\usepackage{rotating}
\usepackage{bbm}
\usepackage[usenames,dvipsnames]{xcolor}  
\usepackage{float}
\usepackage{epsfig}
\usepackage{psfrag}
\usepackage{natbib}
\usepackage{bm}
\usepackage{eucal}
\usepackage{mathrsfs}
\usepackage{braket}
\usepackage{enumerate}
\usepackage{longtable}
\usepackage{subfigure}
\usepackage{bm}
\usepackage{hyperref}
\usepackage{amsfonts}
\usepackage{dcolumn}
\usepackage{bm}
\usepackage{subfigure}

\newcommand{\be}{\begin{equation}}
\newcommand{\ee}{\end{equation}}
\newcommand{\bn}{\begin{eqnarray}}
\newcommand{\en}{\end{eqnarray}}

\usepackage{color} 

\usepackage{hyperref}
\hypersetup{
colorlinks=true,final=true,
        linkcolor=red,
        citecolor=blue,
        filecolor=blue,
        urlcolor=blue,
}
\begin{document}
\title{A DMFT+CTQMC Investigation of Strange Metallicity in Local Quantum Critical Scenario}

\author{Swagata Acharya$^{1}$}\email{swagata@phy.iitkgp.ernet.in}
\author{M. S. Laad$^{2}$}\email{mslaad@imsc.res.in}
\author{A. Taraphder$^{1,3}$}\email{arghya@phy.iitkgp.ernet.in}
\affiliation{$^{1}$Department of Physics, Indian Institute of Technology,
Kharagpur, Kharagpur 721302, India.}
\affiliation{$^{2}$Institute of Mathematical Sciences, Taramani, Chennai 600113, India
}
\affiliation{$^{3}$Centre for Theoretical Studies, Indian Institute of
Technology Kharagpur, Kharagpur 721302, India.}

\begin{abstract}
``Strange'' metallicity is now a pseudonym for a novel metallic state exhibiting anomalous infra-red
(branch-cut) continuum features in one- and two-particle responses.
Here, we employ dynamical mean-field theory (DMFT) using very low-temperature
continuous-time-quantum Monte-Carlo (CTQMC) solver for an extended periodic Anderson model (EPAM) model
to investigate unusual magnetic fluctuations in the strange metal.
 We show how extinction of Landau quasiparticles in the orbital selective Mott phase (OSMP) leads to
$(i)$ qualitative explication of strange transport features and $(ii)$ anomalous quantum critical 
magnetic fluctuations due to critical liquid-like features in dynamical spin fluctuations, 
in excellent accord with data in some $f$-electron systems.
\end{abstract}
\maketitle
Finding the exotic strange metal phase in cuprates and certain $f$-electron systems is 
by now a fundamental problem. 
Strong electronic correlations change the analytic 
structure of the charge- and spin-fluctuation propagators from an infra-red (Landau overdamped at certain critical wave-vectors)
 pole to branch-cut structure. Experimentally, for $f$-electron systems, proximity to antiferromagnetic quantum critical point (AF-QCP) 
seems to be necessary for generating the soft 
electronic ``glue' that results in high $T_{c}$ super conductivity (HTSC). 
Whether the critical electronic glue responsible for Cooper pairing that 
show unconventional superconductivity on the border of $T=0$ destruction of 
anti-ferromagnetic order is linked to AF critical fluctuations or to dualistic 
fluctuations accompanying Kondo breakdown~\cite{pepin} is thus an open issue.  Rephrased somewhat, the issue is: what is the microscopic origin of the anomalous fluctuation spectra in strange metals?  How are we to understand their links to equally anomalous
fermiology and transport in the same picture?  One often sees unique signatures of strange metallicity 
with fractional exponent and $\omega/T$-scaling in inelastic neutron scattering (INS)~\cite{schroder}, 
almost $T$-independent spin relaxation rate in nuclear magnetic resonance (NMR)~\cite{aeppli}, 
quasilinear-in-$T$ resistivity without saturation and unusual low-energy optical response, 
$\simeq \omega^{-\eta}$ with $0.7\le\eta <1.23$~\cite{akrap} in cuprates as well as the 122-Fe arsenides. Theoretical rationalizations of these findings contrast strongly, ranging from local Kondo destruction, to quantum phase transitions (QPTs) associated with proximity to various ($T=0$) ordered or momentum-selective Mott phases.  These may not be mutually exclusive scenarios, but it must be emphasized that specific local quantum critical signatures mentioned above, which {\it require} negligible momentum dependence of the self-energy and two-particle vertices, do not easily fit into traditional itinerant Hertz-Moriya-Millis views of quantum criticality.     

  These issues motivate this work.  
   Here, we show how extinction of Landau Fermi Liquid (LFL) quasiparticles in an OSMP also enables a natural understanding of the unique branch-cut continuum form of
spin- and charge fluctuations solely as a 
consequence of the dualistic (itinerant-localized) character of carriers.  
An OSMP is required, both in the $FL^{*}$ theory~\cite{sachdev} as well as in the Kondo destruction view, both of which describe the same physical picture of a correlated LFL phase separated from the ordered AF phase by a spin liquid region, in which the local moments decouple from the itinerant carriers.  An 
underlying OSM transition in a microscopic fermionic theory can naturally lead to such a situation.  
 We begin with the EPAM~\cite{pepin,laad} $H=H_{band} + H_{int} + H_{hyb}$, where
$H_{band}=\sum_{k,\sigma}\epsilon_{k}c_{k,\sigma}^{\dag}c_{k,\sigma} + \sum_{<i,j>}t_{ff}f_{i,\sigma}^{\dag}f_{j,\sigma}$, $H_{int}=U_{ff}\sum_{i}n_{if\uparrow}n_{if\downarrow} + U_{cc}\sum_{i}n_{ic\uparrow}n_{ic\downarrow} + \sum_{i,\sigma,\sigma'}U_{fc\sigma\sigma'}n_{if}n_{ic}$ and $H_{hyb}=\sum_{k,\sigma}V_{fc}(k)(f_{k,\sigma}^{\dag}c_{k,\sigma}+h.c)$.  

Here $\epsilon_{k},t_{ff}$ are kinetic energies, $U_{cc}, U_{ff}, U_{fc}$ are local Hubbard interactions (intra and inter orbital) and
$V_{fc}$ is inter-orbital hybridization terms for $c$ and $f$ electrons respectively. 
Such a model is a natural one for describing mixed valent $f$-electron systems such as Yb- and Ce-based intermetallics. 
Theoretically, fluctuating valence and the possibility of associated 
soft valence fluctuations, along with its interplay with spin fluctuations are germane issues for study, but 
related aspects have received less attention in literature.
\vspace{0.0em}
\begin{figure}[h!]
\centering
\subfigure[]{\label{f:C11}\epsfig{file=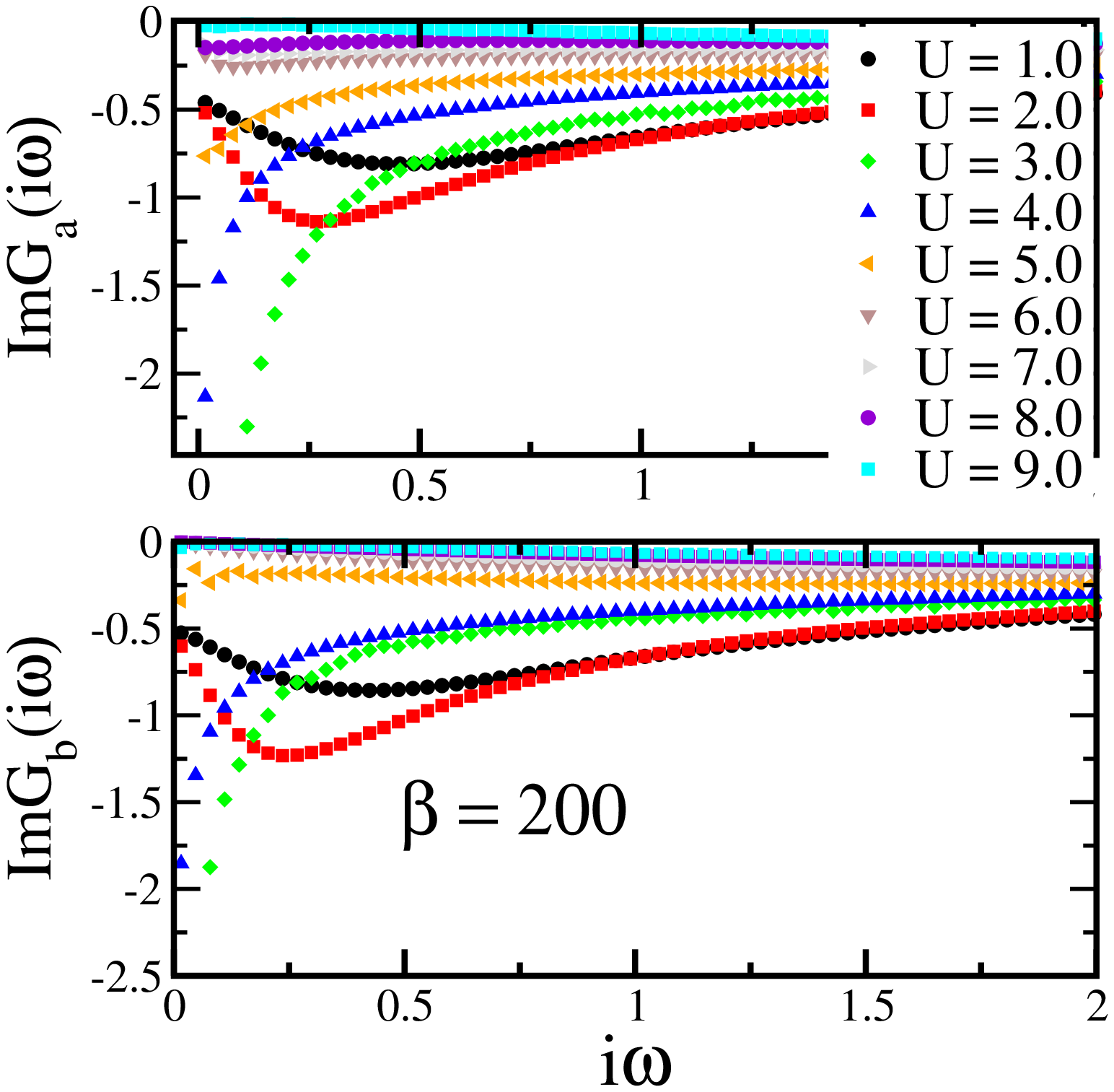,trim=0in 0in 0in 0.0in,
clip=true,width=0.49\linewidth}}\hspace{-0.0\linewidth}
\subfigure[]{\label{f:C21}\epsfig{file=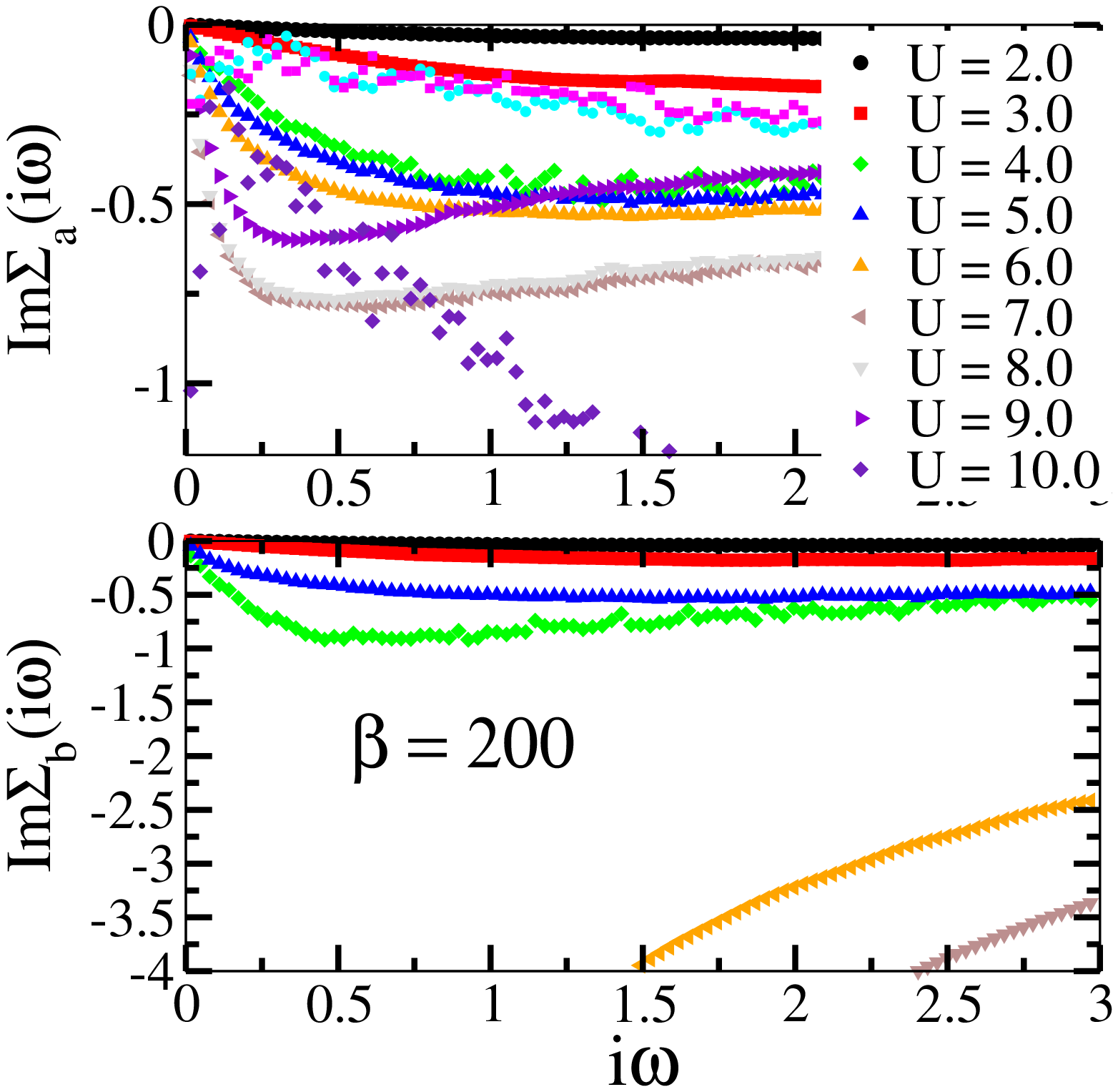,trim=0in 0in 0in 0.0in,
clip=true,width=0.49\linewidth}}\hspace{-0.0\linewidth}
\caption{$ImG_{a,b}$$(\omega)$ and Im$\Sigma_{a,b}(i\omega_{n})$ as a function of 
$i\omega$ for a two-orbital square lattice with inter orbital hybridization $t_{ab}$ (witha d-wave form factor) 
are shown as a function of $U$.}
\label{fig1}
\end{figure}
  It is natural to work in the rotated $a,b$ fermion basis (linear combinations of $c,f$ fermionic operators) which diagonalizes the non-interacting part of $H$ above. 
For $U_{fc}<U_{fc}^{(c)}$ a stable strongly correlated LFL phase gives way, for $U_{fc}>U_{fc}^{(c)}$, to a selective-Mott metal with an infra-red branch-cut form of the one-electron propagator as a consequence of the strong scattering induced ``inverse'' orthogonality catastrophe (OC)~\cite{laad}, as found earlier using the multi-band iterated perturbation theory as an impurity solver in the DMFT context. 
It is obviously of interest to ask whether these appealing features survive upon use of a more exact impurity solver. 
Here, we have solved $H$ (Eq.1) within DMFT using the continuous-time quantum 
Monte Carlo (CT-QMC) hybridization expansion solver~\cite{alps}. 
A very positive feature of our study is in extending CTQMC results to much lower $T=t_{a}/500$ (upto $\beta=500$) than hitherto done.  
Now, local one- and two-particle dynamical responses can be reliably computed, in contrast to diagrammatic solvers like IPT, where two-particle susceptibilities need reliable knowledge of the fully dynamical but local irreducible vertex.  
This is presently a demanding task, unless they can be argued to be irrelevant, as in large-$N$
approaches to DMFT~\cite{sachdev}.
\vspace{0.0em}
\begin{figure}[h!]
\centering
\subfigure[]{\label{f:C11}\epsfig{file=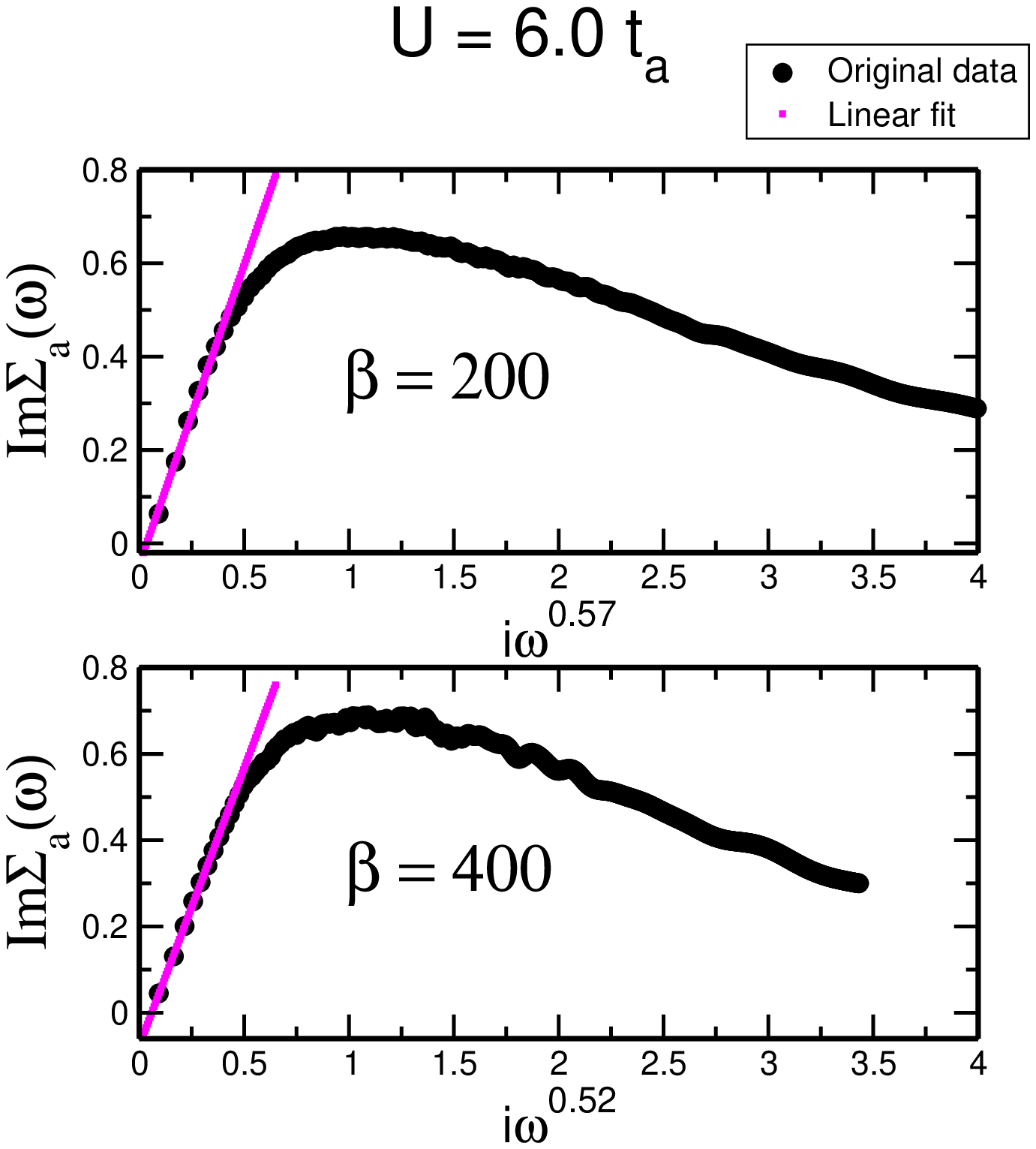,trim=0in 0in 0in 0.0in,
clip=true,width=0.48\linewidth}}\hspace{-0.0\linewidth}
\subfigure[]{\label{f:C21}\epsfig{file=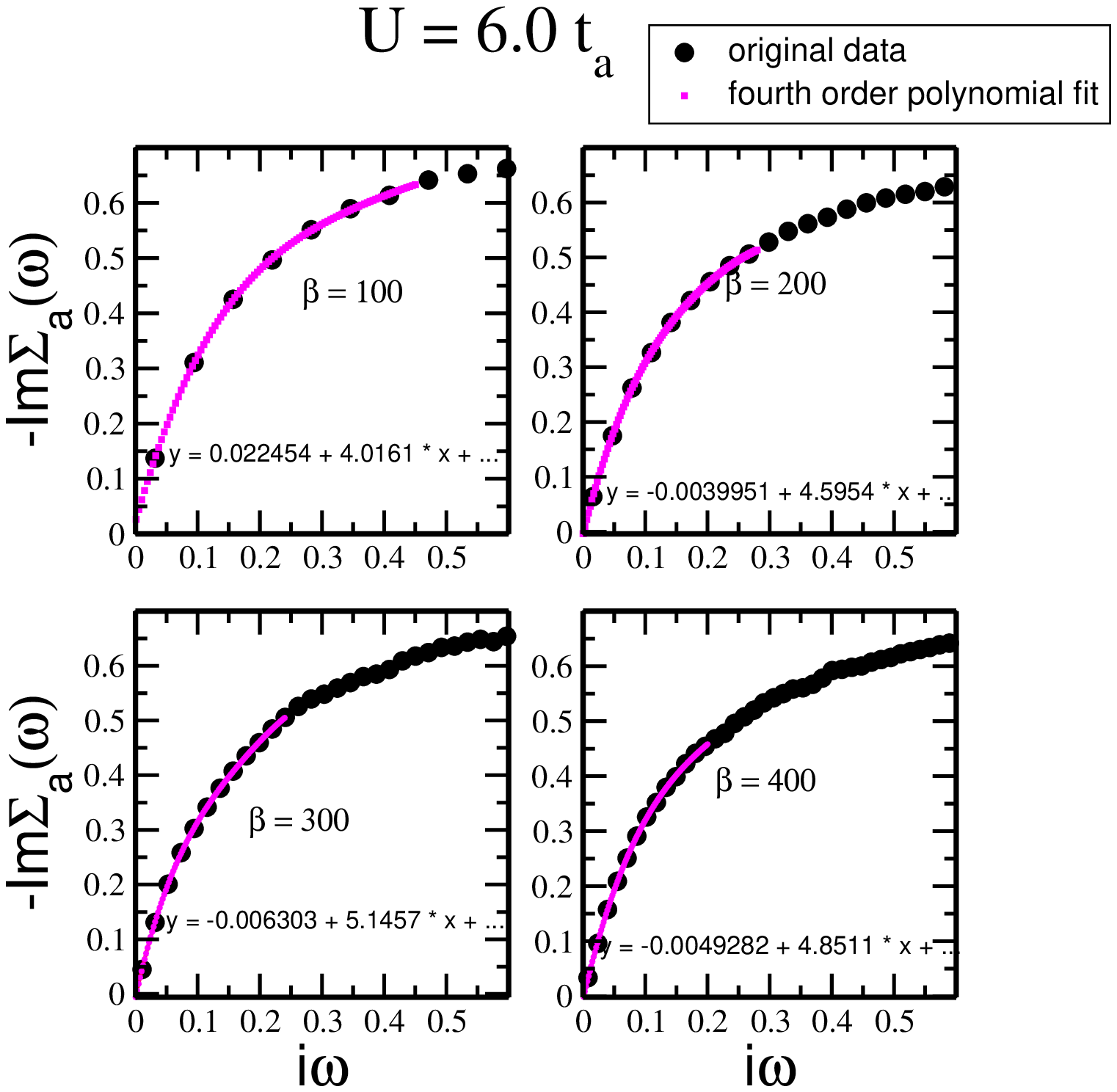,trim=0in 0in 0in 0.0in,
clip=true,width=0.48\linewidth}}\hspace{-0.0\linewidth}
\caption{$Im\Sigma_{a,b}(i\omega)$ as a function of $i\omega$ for a two-orbital square lattice with inter orbital hybridization $t_{ab}$ and $U=6.0$ at low $T$.  Clear power-law behavior in Im$\Sigma_{a}(i\omega)=C(i\omega)^{1-\eta}$ with $\eta=0.43, 0.48$ for $\beta=(k_{B}T)^{-1}=200,400$ up to high energy $O(0.5)$~eV, is visible.  This is a specific characterization of strange metallicity.}
\label{fig2}
\end{figure}

  In the present work, we obtain the full vertex-corrected two-particle dynamical responses for the EPAM using the hybridization expansion-based CTQMC method as an impurity solver. 
Thereafter, we explicitly show that the spin and charge susceptibilities possess a long-time (in imaginary-time) 
quantum-critical form, and use a trick inspired by conformal 
symmetry to extract an effective analytic form in {\it real} frequencies. 
Such a trick has been widely employed in other contexts~\cite{sachdev}. 
This will allow us to discuss transport and magnetic fluctuations in the strange metal phase, and to discuss underlying microscopic features leading to the specific anomalies, in detail.  

  In Fig.~\ref{fig1}, we show the imaginary parts of the one-fermion Green functions for the 
$a,b$ orbital states, along with the corresponding self-energies. 
It is clear thereby that as $U$ and $(U'=U_{ab}=U/3)$ increase, 
the semimetallic behavior for $U=0$ (due to the ${\bf k}$-dependent hybridization) 
changes into a normal, albeit correlated LFL metallicity at 
moderate $U$. 
Beyond $U=4.0$, however, an orbital-selective differentiation of electronic states clearly reveals itself 
in both Im$G_{a,b}(i\omega_{n})$ and Im$\Sigma_{a,b}(i\omega_{n})$: the $a$-states are in an 
incoherent metal state, while the $b$-states are in the Mott insulating regime. 
This dualistic feature exists in the whole parameter region $4.0<U<9.0$ for the given set of 
one-electron parameters in the EPAM, and thus this is an orbital-selective Mott {\it phase}, rather than a QC point. 
For $U>10.0$, we find that a full Mott insulator phase obtains.  
\vspace{0.1em}
\begin{figure}[h!]
\centering
\subfigure[]{\label{f:C11}\epsfig{file=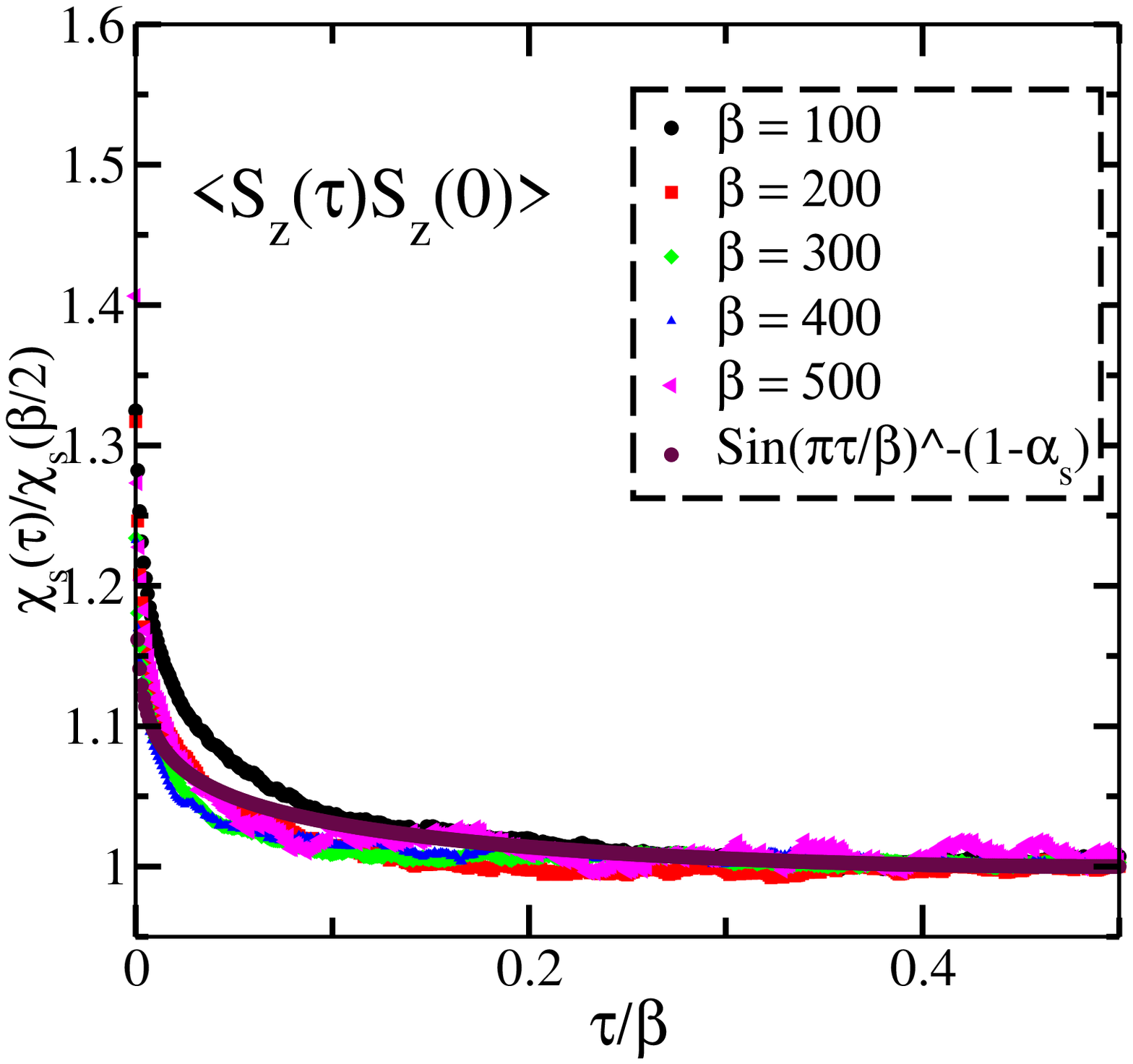,trim=0in 0in 0in 0.0in,
clip=true,width=0.49\linewidth}}\hspace{-0.0\linewidth}
\subfigure[]{\label{f:C21}\epsfig{file=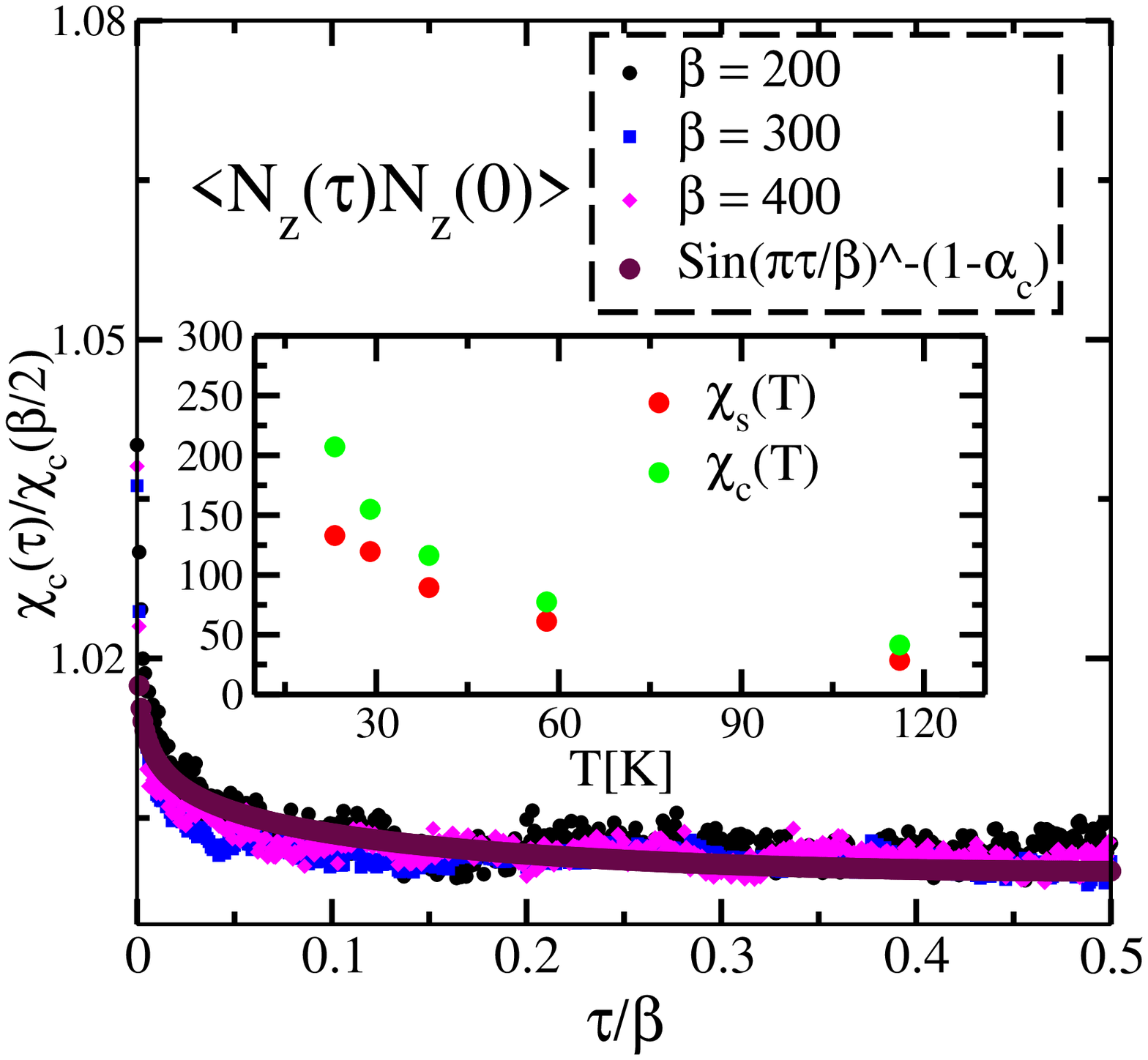,trim=0in 0in 0in 0.0in,
clip=true,width=0.49\linewidth}}\hspace{-0.0\linewidth}
\caption{$Im\chi_{s,c}(\tau)$ as a function of $\tau/\beta$. 
Both, the dynamical spin (s) and charge (c) susceptibilities show 
quantum critical scaling, with $\chi_{s,c}(\tau)/\chi_{s,c}(\beta/2) \simeq [$sin$(\pi\tau/\beta)]^-{(1-\alpha_{s,c})}$,
with $\alpha_{s}(0.9)\neq \alpha_{c}(0.98)$.
Within our numerical accuracy, this constitutes a high-$D$ realization of ``spin-charge separation''.
The inset shows the static local
susceptibilities where $\chi_{c}(T)$ is more singular than $\chi_{s}(T)$ at low temperatures.}
\label{fig3}
\end{figure}

  Focussing on the OSMP, we now exhibit the $a,b$-fermion self-energy, 
Im$\Sigma_{a,b}(i\omega_{n})$, and the full
{\it local} spin susceptibility, Im$\chi_{s}(i\omega_{n})$, 
for the orbital-selective Mott phase Fig.~\ref{fig1},~\ref{fig2}. 
Interestingly, we find clear evidence of anomalous fractional power-law exponents in both. 
In Fig.~\ref{fig2}, we see
that $Im\Sigma_{a}(i\omega_{n})\simeq (i\omega_{n})^{1-\eta}$, 
with $\eta=0.43 (\beta=200)$ and $\eta=0.48 (\beta=400)$. 
In fact, this behavior extends up to rather high energies $O(0.5)$~eV, which is 
characteristic of a multi-particle electronic continuum expected to extend to high energies in a quantum critical regime.  
Remarkably, the {\it local} part
of the dynamical spin susceptibility also exhibits infra-red singular and fractional power-law scaling behavior, characteristic
of the strange metal. We see this as follows. Using the DMFT(CTQMC) results, 
we see that all $\chi_{s}(\tau)$ curves for $U>4.0$ fall on each other and, 
upon a careful fitting with a scaling function characteristic for quantum criticality, 
we find that $\chi_{s}(\tau)/\chi_{s}(\beta/2)$ scales as $[$sin$(\pi\tau/\beta)]^-{(1-\alpha_{s})}$ with $\eta=0.9$ Fig.~\ref{fig3}. This is precisely the scaling form dictated by conformal invariance in the quantum critical region.  It exactly corresponds to the scaling form

\be
\chi_{s}''(\omega) \simeq \omega^{-\eta}F_{s}(\omega/T)
\ee
with $F_{s}(x)=x^{\eta}|\Gamma(\frac{1-\eta}{2}+i\frac{x}{2\pi})|^{2}$sinh$(x/2)$, whence we immediately read off that $T^{-\eta}\chi_{s}''(\omega)$ is a universal scaling function of $\omega/T$, with a fractional-power-law dependence.  Such a unique form for the dynamical spin susceptibility near magnetic QCPs is experimentally seen in more than a few systems by now, specifically in (not obviously proximate to magnetic order), CeCu$_{6-x}$Au$_{x}$~\cite{schroder}, among others. 
On a theoretical level, extinction of Landau fermionic quasiparticles in $G_{b}(k,\omega)$ in the OSMP directly manifests in the emergence of a critical branch-cut continuum in (one-spin-flip) spin-fluctuations. 
Even more interestingly, the dynamical {\it charge} susceptibility also exhibits similar scaling form, but with an 
exponent $\alpha_{c} \neq \alpha_{s}$, implying that spin and charge fluctuations propagate with distinct velocities. 
This is our central result, and we show that it leads to very good accord with 
the unusual normal-state magnetic fluctuation spectra {\it and} transport in 
strange metals. 
Physically, these features emerge as a direct manifestation of emergent, critical ``pseudoparticles'' driving the extinction of stable Landau-damped FL-like collective modes  in the strange metal. 
This is simply because charge and spin fluctuations are themselves constructed from the now incoherent multiparticle continuum, rather than usual Landau quasiparticles.  
\vspace{0.1cm}

  What is the underlying physical origin of these emergent anomalous features?
  As in FL$^{*}$ or Kondo breakdown theories, emergence of critical liquid-like features do {\it not} need a one-to-one association with proximity to $T=0$ magnetic order, since DMFT only accesses strong {\it local} dynamical correlations.  In the OSMP, $V_{fc}(k)$ plays an especially distinct role.
    Explicitly, the tendency of $V_{fc}$ to transfer a $a$-fermion into an $b$-fermion band is dynamically blocked in the OSMT. 
This is because the lower-Hubbard band now 
corresponds to all singly occupied $b$-states, so action of $V_{fc}$ must create a doubly occupied  
(two opposite-spin electrons in the $b$-orbital) intermediate state. 
However, this lies in the {\it upper}
Hubbard band in the $b$-sector and thus the resulting term now has the form 
$V_{fc}'(n_{i,b,-\sigma}b_{i,\sigma}^{\dag}a_{j,\sigma}+h.c)$, which couples the $a$-fermion to a Gutzwiller-projected $b$-fermion,
and thus has no interpretation in terms of a coherent one-electron-like state any more. 
In this sector, this is a high-energy state, and is asymptotically projected out from the low-energy Hilbert space 
(space of states with energy less than the selective-Mott gap). 
It is this emergent projective aspect that is at the root of irrelevance of $V_{fc}(k)$ at one-electron level 
(thus, this is a Kondo destruction~\cite{pepin}) and emergence of strange metal features we find above. 
Thus, we find a mechanism for the destruction of the LFL picture to be very closely related to the 
``hidden-FL'' view of Anderson~\cite{pwa}. 
In absence of a lattice Kondo scale, the hopping of an $b$-electron from site $i$ to the bath or vice-versa 
(and similarly for the $a$-electron) thus
creates a ``suddenly switched-on'' local potential for the $b$-electrons.  
A direct upshot is that the inter-band
spin fluctuations (now local triplet excitons of the type $(a_{i\sigma}^{\dag}b_{i\sigma'}+h.c)$ with $\sigma'=-\sigma$) 
thus also experience a local potential that is (incoherently) switched on and off as a function of time. 
The resulting problem is precisely the ``inverse'' of the Anderson-Nozieres-de Dominicis
X-ray edge problem in the local limit of DMFT, and is a central feature of Anderson's hidden-FL theory~\cite{pwa}.
At two-particle level, this reflects itself in a divergent number of soft, {\it local} spin fluctuation modes, manifesting itself 
as an infra-red singularity along with ``local quantum critical'' $\omega/T$ scaling and anomalous exponents in the spin fluctuation spectrum.  This is completely borne out by our CTQMC results (Fig.~\ref{fig2})  
  The resulting spin polarizability, 
$\Pi(\omega) \simeq \chi_{zz}^{-1}(\omega) \simeq A|\omega|^{1-\alpha}$ in the QC metal,
in stark contrast to $\Pi({\bf q},\omega) \simeq -ia\omega$, known in the heavy
 LFL metal.
The fully renormalized $\chi({\bf q},\omega)$ now reads

\be
\chi({\bf q},\omega)=\frac{1}{\Pi(\omega)+J({\bf q})}
\ee
where $J({\bf q})\simeq -c+({\bf q}-{\bf Q})^{2}+...$ (e.g, $Q=(\pi,\pi)$ for Neel AFM).  At finite $T$,  $\Pi(\omega,T)=AT^{\alpha}f(\omega/T)$.

\vspace{0.4cm}

{\bf Physical Responses In The Strange Metal}

Remarkably, it turns out that our results afford a consistent and surprisingly good quantitative description 
of both transport and neutron results in the strange metal.  Let us consider the consequences of our numerical findings in more detail in this section, with a specific focus on $f$-electron metallic systems.

\noindent $(i)$  The dynamical spin susceptibility shows an explicit $\omega/T$-scaling with anomalous fractional exponent $\alpha_{s}$.  This form is well-documented as the characteristic behavior in real systems where 
quantum critical features cannot be rationalized within a conventional picture based on a quantum Landau-Ginzburg theory.  It is qualitatively consistent with observations in CeCu$_{6-x}$Au$_{x}$~\cite{schroder},
and may also be applicable to other $f$-electron systems where there is independent evidence of the relevance of strong, quasilocal liquid-like critical fluctuations near destruction of magnetic order.  From the form of the dynamical spin susceptibility, we can also immediately read off that the spin relaxation rate in nuclear-magnetic resonance (NMR) studies will vary very weakly with $T$, as $1/T_{1}\simeq T^{0.1}$ or that 
$1/T_{1}T \simeq T^{-0.9}$.  This is quite close to the measured $1/T_{1}T \simeq T^{-1}$ in near-optimally
doped cuprates~\cite{aeppli}, and it will be of interest to study NMR relaxation rates near the QCPs in $f$-electron systems to see whether these local critical features are more generic.

\noindent $(i)$  The fact that Im$\Sigma_{a,b}(i\omega)\simeq (i\omega)^{1-\eta}$ with $(1-\eta)=0.52,0.57$ for $\beta=400,200$ allows study of $dc$ and $ac$ conductivities without any further approximation within DMFT. 
This is because irreducible vertex corrections to the current correlation functions for the conductivities in the Bethe-Salpeter equation rigorously vanish in this limit.
Transport is now entirely determined by the ``simple'' bubble diagram composed of the full DMFT one-electron propagators. 
Explicitly, the $dc$ resistivity is now $\rho_{dc}(T)\simeq T^{2(1-\eta}=T^{1.04}$ for 
$\beta=400$ and $T^{1.14}$ for $\beta=200$.  Such an unusual resistivity, $\simeq T$, has long been known as one
of the defining features of the strange metal, not only in cuprates, but also in $D=3$ $f$-electron systems.
In YbRh$_{2}$Si$_{2}$ (YRS), for example, the $dc$ resistivity does vary like $T^{n}$ with $n$ slightly higher than unity~\cite{steglich}.  The optical conductivity can also be readily estimated as $\sigma(\omega)\simeq \omega^{-2(1-\eta)}=\omega^{-1.04}, T^{-1.14}$. 
This is indeed consistent with data in underdoped cuprates as well as in the $D=3$ 122-Fe arsenides~\cite{akrap}. 
OSMP are known to be generic in Fe arsenides by now~\cite{capone},
and our work now explicitly links the anomalous optical response, at least in 122-Fe arsenides, to an underlying
OSMP. 
It would be very interesting to search for such manifestations of 
anomalous metallicity in $f$-electron systems, 
in particular in YRS, where limited evidence of a 
quasi-linear-in-$\omega$ scattering rate and anomalous increase in optical 
effective mass at low energy indeed exists. 
Another promising avenue is that charge fluctuations seem more 
singular than the spin fluctuations (inset of Fig.~\ref{fig3}), 
so one would expect a valence transition to occur prior to onset of magnetism. 
We leave these issues for future study.

\vspace{0.1cm}
{\bf CONCLUSION}
\vspace{0.1cm}

   To conclude, we have presented strong numerical evidence for anomalous spin and charge 
fluctuation spectra emerging in an orbital-selective Mott phase in an extended periodic Anderson model. 
This is the basic model for
a wide variety of mixed-valent $f$-electron systems, 
and is also suited to describe multi-orbital transition-metal 
oxides in a different parameter space. 
We detail how extinction of the Landau quasiparticle pole structure in the 
one-electron propagators in the OSMP leads to such a critical fluctuation spectrum, and point 
out that these features are qualitatively consistent with characteristic signatures of strange metallicity
in mixed-valent rare-earth systems. 
We expect these results to have broader applicability.
\bibliographystyle{apsrev4-1}

\end{document}